# Globally stable microresonator Turing pattern formation for coherent high-power THz radiation on-chip

Shu-Wei Huang[1,*], Jinghui Yang[1,*], Shang-Hua Yang[2], Mingbin Yu[3], Dim-Lee Kwong[3], T. Zelevinsky[4], Mona Jarrahi[2], and Chee Wei Wong[1,*]

[1] Fang Lu Mesoscopic Optics and Quantum Electronics Laboratory, University of California, Los Angeles, CA 90095, USA.

[2] Terahertz Electronics Laboratory, University of California, Los Angeles, CA 90095, USA.

[3] Institute of Microelectronics, A*STAR, Singapore 117865, Singapore.

[4] Department of Physics, Columbia University, New York, NY 10027, USA.

* Author e-mails: swhuang@seas.ucla.edu; yangjh@seas.ucla.edu; cheewei.wong@ucla.edu

**In nonlinear microresonators driven by continuous-wave (cw) lasers, Turing patterns have been studied in the formalism of Lugiato-Lefever equation with emphasis on its high coherence and exceptional robustness against perturbations. Destabilization of Turing pattern and transition to spatio-temporal chaos, however, limits the available energy carried in the Turing rolls and prevents further harvest of their high coherence and robustness to noise. Here we report a novel scheme to circumvent such destabilization, by incorporating the effect of local mode hybridizations, and attain globally stable Turing pattern formation in chip-scale nonlinear oscillators, achieving a record high power conversion efficiency of 45% and an elevated peak-to-valley contrast of 100. The stationary Turing pattern is discretely tunable across 430 GHz on a THz carrier, with a fractional frequency sideband non-uniformity measured at $7.3 \times 10^{-14}$. We demonstrate the simultaneous microwave and optical coherence of the Turing rolls at different evolution stages through ultrafast optical correlation techniques. The free-running Turing roll coherence, 9 kHz in 200 ms and 160 kHz in 20 minutes, is transferred onto a plasmonic photomixer for one of the highest power THz coherent generation at room-temperature, with 1.1% optical-to-THz power conversion. Its long-term stability can be further improved by more than two orders of magnitude, reaching an Allan deviation of $6 \times 10^{-10}$ at 100 s, with a simple computer-aided slow feedback control. The demonstrated on-chip coherent high-power Turing-THz system is promising to find applications in astrophysics, medical imaging, and wireless communications.**

The spontaneous formation of stationary periodic patterns from homogenous background firstly elucidated by Turing has served as the basis for developmental biology morphogenesis, chemical kinetics far-from-equilibrium, and the formation of fractals and chaos in nonlinear dynamics [1–3]. Of both conceptual importance and practical interest, optical Turing pattern formation has been theoretically proposed and investigated in cw-laser-pumped Kerr-active microresonators [4–7], with emphasis on its high coherence and exceptional robustness against perturbations. Generally the Kerr-active microresoantor is designed to possess anomalous group velocity dispersion (GVD) for convenient phase matching fulfillment and the formation dynamics of the spontaneous patterns can be described by the Lugiato-Lefever equation [4]. As the driving laser is frequency tuned into the cavity resonance from the blue side, stable Turing rolls first spontaneously emerge from the background, then quickly destabilize into spatio-temporal chaos [8,9], and eventually transition into dissipative Kerr solitons [10–14] or Kerr frequency comb [15–17] (Figure S1). We note that the Turing roll in this dispersion regime, despite its optimally coherent properties [18], only exists in a limited phase space and its quick destabilization into chaos limits the attainable power conversion efficiency (Figures 1a and 1b), preventing further harvest of the high coherence and the noise robustness of the Turing pattern.

To expand the stability zone and attain higher power conversion, we approach the spontaneous Turing pattern formation in a distinctly different way. Our Kerr-active microresonator is designed to possess a large globally normal GVD of 100 fs$^2$/mm (Figure S2) and thus phase matching is strictly forbidden unless local dispersion anomalies are introduced to the system, providing additional GVD to balance the nonlinearity locally. In our high $Q$ microresonator (loaded quality factor of 3.7×10$^5$), such anomalies result from the hybridization of two transverse modes (TM$_{11}$ and TM$_{21}$) with distinct free spectral ranges (FSRs) when their resonant frequencies are in the vicinity of each other [19–24]. The spectral position of the mode hybridization defines the Turing pattern formation dynamics and it can be changed by the design of the FSR difference. As the balance between the GVD and the Kerr nonlinearity is only fulfilled locally in the confined spectral range where mode hybridization occurs, sub-comb growth and subsequent destabilization of the Turing pattern is avoided (Figure 1c). By enabling the deeper driving-into-resonance without transition into chaos or soliton states, the conversion efficiency from pump to Turing pattern can thus be significantly enhanced in our system. The mode hybridization mediated phase matching – by adjusting the relative frequency between the pump mode and the mode hybridization position

– further enables the repetition rate of the Turing roll to be discretely tunable. Moreover, the Turing roll spectra can exhibit controllable asymmetry through registering the pump mode on different sides of the first local-mode hybridization region $\varepsilon_1$ (Figure 2a).

Here we demonstrate the scheme of incorporating the mode hybridization effect to attain globally stable microresonator Turing patterns in chip-scale nonlinear nitride cavities, achieving an unprecedented pump depletion and a record high power conversion efficiency of 45% with an elevated peak-to-valley contrast of 100. We interrogate the commensurate and coherent nature of the spontaneous dissipative structure with ultrafast optical intensity autocorrelation, microwave spectral noise analysis, and heterodyne beating against a benchmark fiber frequency comb. The fractional frequency sideband non-uniformity of the Turing pattern is measured at $7.3\times10^{-14}$, with a short-term (200 ms sweep time) linewidth of 9 kHz and a long-term (over 20 minutes) fluctuation of 160 kHz in the free-running mode. The long-term stability can be further improved by more than two orders of magnitude, reaching an Allan deviation of $6\times10^{-10}$ at 100 s, with a simple computer-aided slow feedback control. Towards THz applications, we then transfer the Turing pattern optical coherence to the THz carrier through a plasmonic ErAs:InGaAs photomixer, generating up to 600 µW THz radiation power at room temperature. The carrier frequency is discretely tunable over a broadband from 1.14 THz to 1.57 THz. The demonstrated coherent high-power Turing-THz system offers the potential to be the room temperature on-chip THz local oscillator for astrophysics, medical imaging, and wireless communication [25–30].

Here a TM-polarized cw laser with an optical power of 29.5 dBm is frequency tuned from the blue side of the cavity resonance to trigger the Turing pattern formation. The measured spectra of the spontaneous Turing patterns generated from ring resonators with different radii are shown in Figure 2a, with the pump illustrated in blue and the mode hybridization positions labeled as red dashed lines. Despite very similar GVD ($\beta_2$) from different radii ring resonators, the Turing patterns show spectral shapes distinct from each other. Specifically the spectral lines on the side of the first mode hybridization position are suppressed due to the increasing phase mismatch associated with the mode hybridization induced local dispersion disruption. With the first sideband pair ($m = \pm 1$) phase matched due to the additional contribution from the mode hybridization on mode 1, then the phase matching condition can be written as:

$$\Delta k(2\omega_0 - \omega_1 - \omega_{-1}) = \beta_2 \Delta^2 + \gamma P_{int} - \varepsilon = 0 \,,$$

where $\varepsilon$ represents the contribution from the mode hybridization, $\Delta$ the Turing roll repetition rate, $\gamma$ the nonlinear Kerr coefficient, and $P_{int}$ the intracavity power. Here $k_1 = k_0 + \beta'\Delta + \frac{\beta_2}{2}\Delta^2 + \varepsilon$, and $k_m = k_0 + \beta'(m\Delta) + \frac{\beta_2}{2}(m\Delta)^2$, where $\beta'$ is the group velocity. Then the phase matching condition for the first cascaded FWM on *either* side of the pump can be written as:

$$\Delta k(2\omega_1 - \omega_2 - \omega_0) = \beta_2 \Delta^2 + \gamma P_{int} + 2\varepsilon = 3\varepsilon$$
$$\Delta k(2\omega_{-1} - \omega_0 - \omega_{-2}) = \beta_2 \Delta^2 + \gamma P_{int} = \varepsilon .$$

The phase mismatch on the side of the mode crossing position is three times larger than the other process and thus the symmetry of the Turing roll spectra is broken. Similar symmetry breaking by mode hybridization has also been demonstrated in the microwave photonics recently [31]. Turing roll repetition rates also show dramatic variations, $\approx 640$ GHz in the 180 μm radius ring and $\approx 1.72$ THz in the 160 μm and 200 μm rings, that cannot be solely explained by the change in the cavity round-trip time. These features are direct consequences of the unique phase matching configuration employed in our design, with critical roles in the efficient coherence transfer from Turing pattern to THz radiation to be detailed subsequently. Of note, the mode hybridization induced local dispersion disruption $\varepsilon$ can be dynamically tuned by changing the temperature of the microresonator, thereby providing an additional dimension to control the Turing pattern formation dynamics (detailed in the Supplemental Materials II).

Next we focus on the results generated from the 160 μm radius ring because of its energy concentration in the wavelength range shorter than 1570 nm, overlapping better with the spectral response of our plasmonic photomixer discussed later. Figure 2b and 2c shows the pump and total transmitted intensities, measured simultaneously for different detunings. The pump transmission shows a triangular tuning curve with a strong dip from normalized unity into $\approx 10\%$ of the original transmission, while total transmission shows only a small drop from 48% to 41% – this provides the evidence of the efficient total energy transfer from the pump into the complete Turing pattern sidebands. To examine this further, Figure 2d plots the corresponding Turing roll spectra at different detuning stages: stable spontaneous Turing pattern are observed without any sign of destabilization at all detunings. The power conversion efficiency, defined as the integrated power of the output Turing roll divided by the on-chip pump power, reaches as high as 45% at the stage III (detailed in the Supplemental Materials III). Closer to resonance at the stage V, an even stronger pump depletion is achieved, with the pump intensity 2-dB lower than even the first modulation sidebands.

We conduct a series of ultrafast optical intensity autocorrelation (IAC) measurements, as shown in Figure 3a, to investigate the temporal structure of the Turing patterns at different evolving stages. At all the stages, stable and strong quasi-sinusoidal oscillations are each observed. While pumping closer to resonance results in monotonic increase in the pump depletion (Figure 2d), counterintuitively the IAC traces show a discernible minimum background between stage II and stage III. As elaborated later, it is a direct consequence of strong pump depletion. At stage III, a peak-to-valley intensity contrast ratio of more than 100 is achieved (Figure S8). We next perform extensive measurements to examine the coherence of the spontaneous Turing formation, illustrated in Figure 3b to 3e. The RF amplitude noise spectra of the Turing pattern up to 3 GHz, six times the cavity linewidth, shows an absence of RF peaks and a noise level at the instrumentation detection background limit, indicative of the existence of a single Turing roll family with commensurate repetition rates (Figure 3b). We further heterodyne beat the Turing sidebands ($m = \pm 1$ and pump) against a benchmark fiber frequency comb (see Appendix) and perform ratio counting of the sidebands to interrogate the frequency uniformity, which sets the fundamental limit on the coherence transfer from the Turing pattern (Figure 3c). When the Turing roll repetition rate is made non-divisible by the fiber frequency comb spacing, the beat frequencies of consecutive sidebands will be an arithmetic sequence. Namely, $\delta_2 = \delta_1 + \Delta = \delta_0 + 2\Delta$. Here we make the common difference, $\Delta$, to be 1 MHz. Ideally, the ratio between $\delta_2 - \delta_0$ and $\delta_1 - \delta_0$, $R$, should be 2 and deviation from this ratio, $\varepsilon_R$, is a measure of the sideband frequency non-uniformity, $\varepsilon = \varepsilon_R \cdot \Delta$. Excellent sideband uniformity of the Turing pattern is observed at all evolving stages with the average non-uniformity measured at 125 mHz, $7.3 \times 10^{-14}$ when referenced to the Turing pattern repetition rate at 1.72 THz (Figure 3c). Figure 3d next shows the self-heterodyne beat note of the first sideband (see Appendix), demonstrating the Turing lines down to a pump-coherence-limited linewidth of 500 kHz. The linewidth measurements independently confirm the good coherence of the Turing rolls at all detunings. Figure 3e also shows the real-time power monitoring of the four strongest sidebands. All sidebands present similar intensity noise of less than 1% (integrated from 100 Hz to 100 MHz) and no cyclic energy exchange between sidebands is observed, excluding the possibility of breathing dynamics with conserved total power and supporting the evidence of stationary Turing pattern formation with extensive stability zone.

To examine whether the temporal shape of the spontaneous Turing pattern is subjected to the perturbation in the initial condition and the pump detuning scan, we performed the IAC measurements at three different tuning speeds and two independent starts. Each of the Turing IAC dynamics remains identical to each other (Figure S9), illustrating the good robustness of the Turing patterns. To understand the dynamics better, ideal IAC traces from transform-limited Turing patterns are superimposed onto the measured IAC traces (Figure 3a; in red). As the Turing pattern is driven closer to resonance, there is increasing discrepancy of the measured pattern from the transform-limit. The change of temporal shapes without the coherence loss implies that the spectral phase of the Turing pattern varies at the different evolving stages and thus different external phase compensation strategies are necessary if the temporal properties of the Turing pattern are to be fully utilized. The spectral phase variation can be understood as the consequence of the pump phase slip around the resonance. The relationship between the output and the intracavity pump power can be written as:

$$A_{p,out} = -\frac{\gamma_c - \gamma_\alpha - i\delta}{\sqrt{2\gamma_c}} \sqrt{T_R} A_{p,cav} ,$$

where $\gamma_c$ and $\gamma_\alpha$ are the half-width half-maximum linewidths associated with the coupling losses and the intrinsic cavity losses respectively. In our microresonator, $\gamma_c = 160$ MHz and $\gamma_\alpha = 90$ MHz. The output pump will experience a $\pi$ phase shift as it traverses through the resonance (Figure S9a). Such phase slip is due to the interference between the intracavity and the input pump and thus the other sidebands of the Turing roll will not experience such a phase shift. This additional pump phase slip results in the observed change of temporal structure. The distinct responses to dispersion (illustrated in Figure S10; aiding between different Turing states) are well captured by considering the pump phase offset (Figure S11b). Of note, most of the phase slip happens very close to cavity resonance and thus the observation of its effect is attributed to the unique design of our microresonator, which utilizes the local mode hybridization to fulfill the phase matching of spontaneous Turing pattern formation and greatly suppresses the Turing pattern destabilization even when the pump is deep into the resonance.

The robustness, tunability, good coherence, and high efficiency of the demonstrated Turing roll make it an excellent photomixer pump for narrow linewidth tunable THz radiation. Different from deriving the pump from a mode-locked laser [29], the Turing pattern offers the advantage of efficient power use and reduced system complexity as its quasi-sinusoidal intensity profile (Figure S8) is directly applicable as a photomixer pump. Other demonstrated photomixer pump source until now [32] include independent lasers with frequency stabilization [33,34], single laser with active high-speed phase modulation [30] and tunable dual-mode lasers [35–39]. While phase-locking two independent lasers to external frequency references can potentially provide the ultimate coherence, it suffers from greatly increased system complexity. Using a single laser with active phase modulation simplifies the system, but high-speed phase modulation with a bandwidth higher than 1 THz is technologically challenging. Tunable dual-mode laser is attractive in its compact footprint and broadband tunability, but it has a large long-term frequency drift due to the uncommon paths taken by the two modes and solving it again requires a sophisticated phase-locking technique. The demonstrated Turing roll provides yet another promising option as it not only is intrinsically compatible with high power operation but also offers a balance between highest coherence and lowest system complexity.

To characterize the stability of the free-running Turing roll repetition rate, which determines the linewidth and fluctuation of the THz radiation, we beat the pump and one of the sideband with the adjacent fiber laser frequency comb lines and electrically mix the two signals to get the beat note at the frequency difference as shown in Figure S13. Figure 4 shows the long-term frequency fluctuation, measuring a root-mean-square frequency fluctuation of 160 kHz over 20 minutes. The left inset shows the linewidth of the beat note with a sweep time of 200 ms, measuring a narrow FWHM linewidth of 9 kHz; the right inset shows the frequency stability of the beat note, measuring an Allan deviation of $1.15 \times 10^{-8} \cdot \sqrt{\tau}$ when referenced to the THz carrier. Its long-term stability can be further improved by a cost-effective computer-aided slow feedback control of the Turing sideband power. The feedback bandwidth is about 1 Hz. As the microresonator is housed in a temperature-controlled enclosure, we attribute the instability of the Turing roll repetition rate mainly to the effects associated with the fluctuation of intra-cavity power, which at the same time has major impact on the Turing sideband power. Thus by stabilizing the Turing sideband power, the fluctuation of intra-cavity power is reduced and consequently the long-term stability of Turing roll repetition rate is improved (Figure S16). When the feedback loop is engaged, the frequency stability is greatly improved as shown in Figure 5a. Figure 5b shows the Allan deviation after stabilization, measuring an improvement of more than two orders of magnitude and reaching $6 \times 10^{-10}$ at 100 s with an inverse linear dependence on the gate time.

To convert the Turing pattern into the THz radiation, we fabricate an ErAs:InGaAs plasmonic photomixer (Figure 6a) which features a good spectral response from 0.8 to 1.6 THz due to the logarithmic spiral antenna design [40]. A C/L WDM filter, followed by an erbium doped fiber amplifier (EDFA), is used to selectively amplify the Turing pattern sidebands in the 1530 to 1565 nm C-band range. Of note, the EDFA is necessary only because our current $Si_3N_4$ microresonator has a strong $Q$-factor roll-off in C-band and thus we are limited to pump it in the L-band. Figure 6b shows four examples illustrating the tunability of the Turing roll repetition rate (1.14 to 1.57 THz), which in turn determines the THz frequency, by tuning the chip temperature or the pump wavelength. Figure 6c plots the room-temperature radiated THz power as a function of the optical pump power, showing a nearly quadratic dependence even at the maximum available pump power of 54 mW, without much saturation roll-off in the THz generation. Up to 600 µW THz radiation power is generated with an optical-to-THz power conversion efficiency of 1.1%.

In summary, we demonstrate the scheme of incorporating the mode hybridization effect to attain globally stable microresonator Turing patterns in chip-scale nonlinear nitride cavities, achieving an unprecedented pump depletion and a record high power conversion efficiency of 45% with an elevated peak-to-valley contrast of 100. Controllable asymmetry in the Turing roll spectrum is also demonstrated. The fractional frequency sideband non-uniformity of the Turing pattern is measured down to $7.3 \times 10^{-14}$, with a short-term (200 ms sweep time) linewidth of 9 kHz and a long-term (over 20 minutes) fluctuation of 160 kHz in the free-running mode. The long-term stability can be further improved to sub-kHz with a simple computer-aided slow feedback control. We observe the temporal shapes of the Turing patterns change with respect to pump detuning, not because of the coherence loss but the pump phase slip near the cavity resonance. The robustness, tunability, good coherence, and high efficiency of the demonstrated Turing pattern make it an excellent photomixer pump for narrow linewidth tunable THz radiation. Pumping a novel ErAs:InGaAs plasmonic photomixer, we then transfer the Turing pattern optical coherence to the THz carrier and generate up to 600 µW THz radiation power at room temperature. The carrier frequency is discretely tunable over a broadband from 1.14 THz to 1.57 THz. The demonstrated coherent high-power Turing-THz system offers the potential to be the room temperature on-chip THz local oscillator for astrophysics, medical imaging, and wireless communication

**Acknowledgements:** The authors thank discussions with Bart McGuyer. The authors acknowledge funding support from ONR (N00014-14-1-0041), AFOSR Young Investigator

Award (FA9550-15-1-0081), DARPA (HR0011-15-2-0014), and NIST Precision Measurement Grant (60NANB13D163).

**Appendix A: $Si_3N_4$ microresonator fabrication**

First a 3 μm thick oxide layer is deposited via plasma-enhanced chemical vapor deposition (PECVD) on *p*-type 8" silicon wafers to serve as the under-cladding oxide. Then low-pressure chemical vapor deposition (LPCVD) is used to deposit a 725 nm silicon nitride for the ring resonators, with a gas mixture of $SiH_2Cl_2$ and $NH_3$. The resulting silicon nitride layer is patterned by optimized 248 nm deep-ultraviolet lithography and etched down to the buried oxide layer via optimized reactive ion dry etching. Under a scanning electron microscope, the etched sidewall verticality is measured to be 88º and the residual sidewall imperfections result in the mode hybridization effect described in the main text. Next the silicon nitride ring resonators are over-cladded with a 3 μm thick oxide layer, deposited initially with LPCVD (500 nm) and then with PECVD (2500 nm). The propagation loss of the silicon nitride waveguide is measured to be 0.2 dB/cm at the pump wavelength. Cold cavity properties of the microresonator is precisely characterized by a hydrogen cyanide referenced swept-wavelength interferometer (Supplemental Materials VI).

**Appendix B: Ultrafast intensity autocorrelation and self-heterodyne characterization**

The optical intensity autocorrelator consists of a 1 mm thick *β*-BBO crystal, supporting a bandwidth of 200 nm, and a silicon avalanche photodiode, supporting a sensitivity of 100 μW. The setup is configured in a non-collinear geometry and careful checks are done before measurements to ensure only background-free second harmonic signals are collected. The use of dispersive optics is also minimized such that the additional dispersion introduced to the pulse is only -50 $fs^2$. To further confirm the coherence of the Turing sidebands, self-heterodyne linewidth measurements are performed with detailed setup shown in Supplemental Materials VI. With a 200 MHz acousto-optic modulator and 24.5 μs delay provided by a 5 km single mode optical fiber, our self-heterodyne setup shows a minimum linewidth at 13 kHz, about 40 times smaller than the pump laser linewidth (≈ 500 kHz).

**Appendix C: Turing pattern uniformity measurements against a fiber frequency comb**

Sideband frequency uniformity of the Turing pattern is measured by heterodyne referencing against a benchmark optical fiber laser frequency comb, Menlo FC1500-250-WG. As detailed in Supplemental Materials VI, the spontaneous Turing roll is split into 3 beat detection units (*m* = ± 1 and pump) with more than 40 dB signal to noise ratio at 100 kHz resolution bandwidth. The sideband beat notes ($\delta_1$ and $\delta_2$) are mixed with the pump beat note ($\delta_0$) to cancel the residual pump frequency instability, with the frequency counter operating in the ratio counting mode to circumvent synchronization of the simultaneous beat note measurements.

**Appendix D: Plasmonic photomixer fabrication**

The plasmonic contact electrode gratings are designed to have 200 nm pitch, 100 nm metal width, 5/45 nm Ti/Au height, and 250 nm thick silicon nitride anti-reflection (AR) coating. They are patterned with electron-beam lithography followed by deposition of Ti/Au and liftoff. A 250 nm silicon nitride AR coating is then deposited with PECVD. Contact vias are patterned with optical lithography and etched via dry plasma etching. Finally, the logarithmic spiral antennas and bias lines are patterned again with optical lithography, followed by deposition of Ti/Au and liftoff. The fabricated plasmonic photomixers are then mounted on a hyper-hemispherical silicon lens to improve the THz radiation collection.

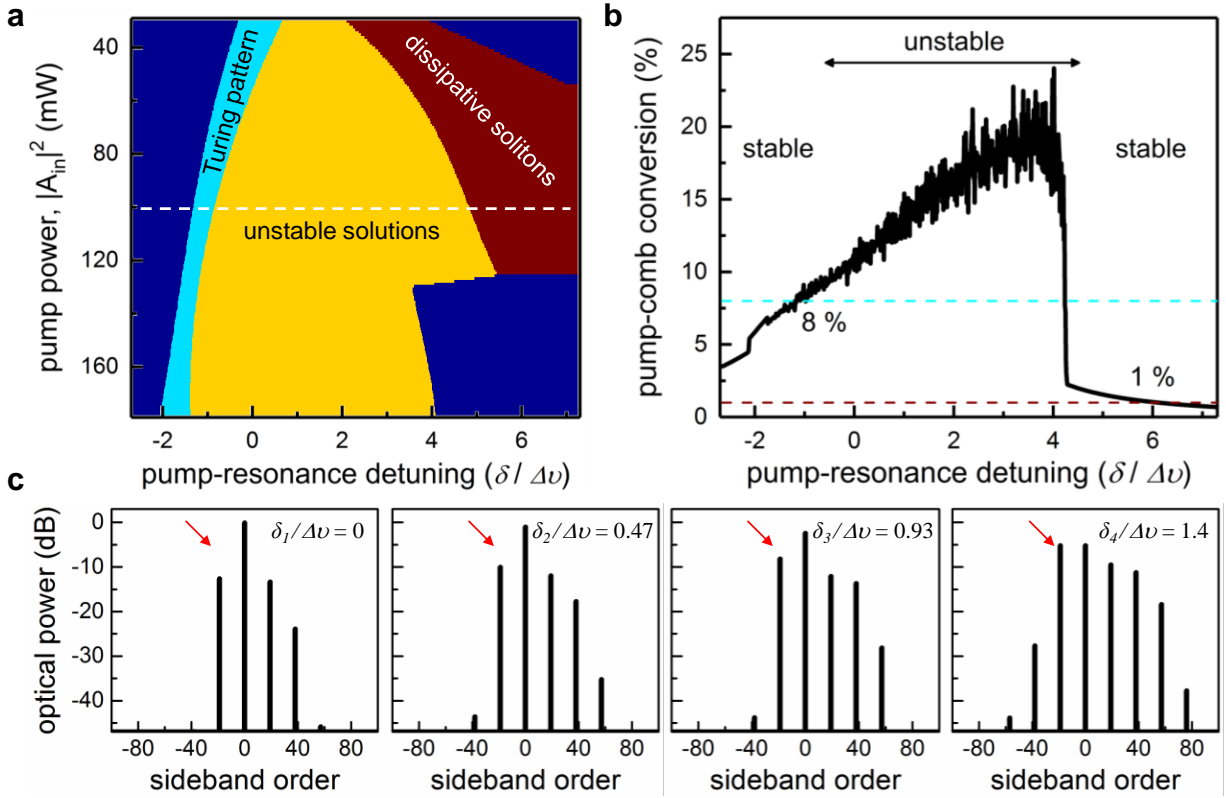

**Figure 1.** (**a**) Stability diagram of the Turing pattern in the anomalous GVD regime. Light blue: region of stable Turing pattern; yellow: region of breathers and spatio-temporal chaos; red: region of soliton and soliton molecules. (**b**) Pump-to-comb power conversion efficiency along the white dashed line in (**a**), showing only 8% maximum power conversions to Turing pattern before destabilization and transition to chaos happens. (**c**) Simulation of the Turing roll in the normal dispersion microresonator (1300 fs$^2$), where the Turing roll is excited by local mode hybridization. The red arrow points to the mode where local dispersion disruption is introduced in the Lugiato-Lefever model. It shows an apparent asymmetry and can be tuned further into resonance without triggering the sub-comb growth and the associated Turing pattern destabilization, expanding the stability zone and opening the route to harness Turing pattern's high coherence and exceptional robustness at high optical powers. Of note, the pump-resonance detuning used in the simulation is referenced to the cold cavity resonance frequency.

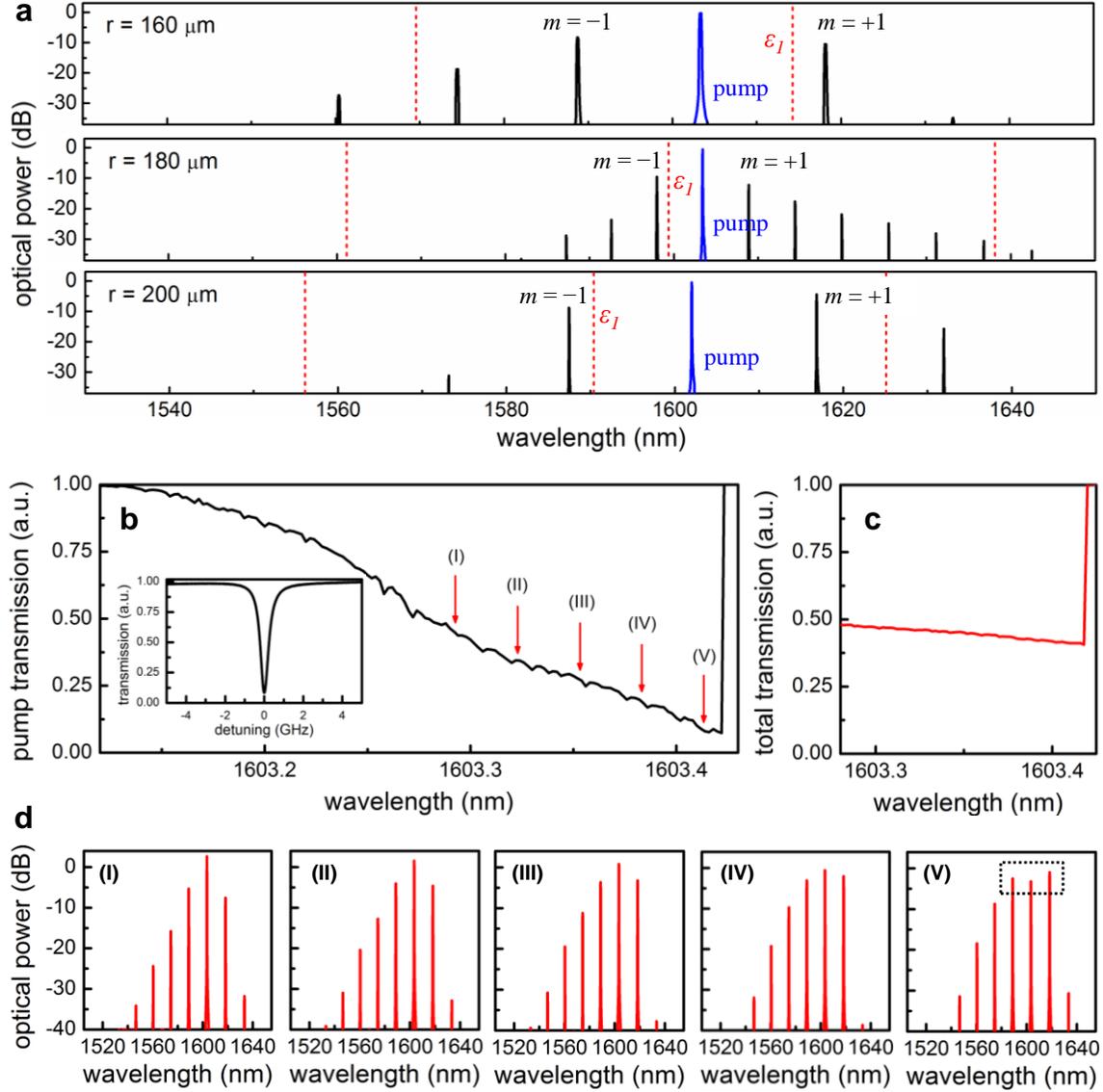

**Figure 2.** (**a**) Turing rolls generated from ring resonators with different radii. Even though the GVD of the ring resonators differ by less than 2 fs$^2$/mm, the TM$_{11}$-TM$_{21}$ mode hybridization positions (red dashed lines) with respect to the pump (blue lines) shift due to the change in the ring radii, resulting in abrupt dispersion variations locally and very different spontaneous Turing patterns. The Turing roll repetition rates are 1.72 THz (12×FSR) for the 160 μm radius ring, 0.64 THz (5×FSR) for the 180 μm radius ring, and 1.72 THz (15×FSR) for the 200 μm radius ring. (**b**) Pump-cavity transmission as a function of the pump wavelength, labeling the detunings where different Turing roll stages are generated. Inset: The cold resonance of the pump mode, measuring a loaded Lorentzian linewidth of 500 MHz and a loaded quality factor of $3.7 \times 10^5$. (**c**) Total cavity transmission as a function of the pump wavelength in the range where Turing roll is generated. Compared to the pump-cavity transmission, the total cavity transmission shows a less apparent decrease as the pump is tuned into the resonance, confirming an efficient power conversion from the pump to the generated Turing lines. (**d**) Example Turing roll spectra at different stages. At stage V, even a highly depleted pump close to the resonance is observed in the measurement, illustrated in the dashed box.

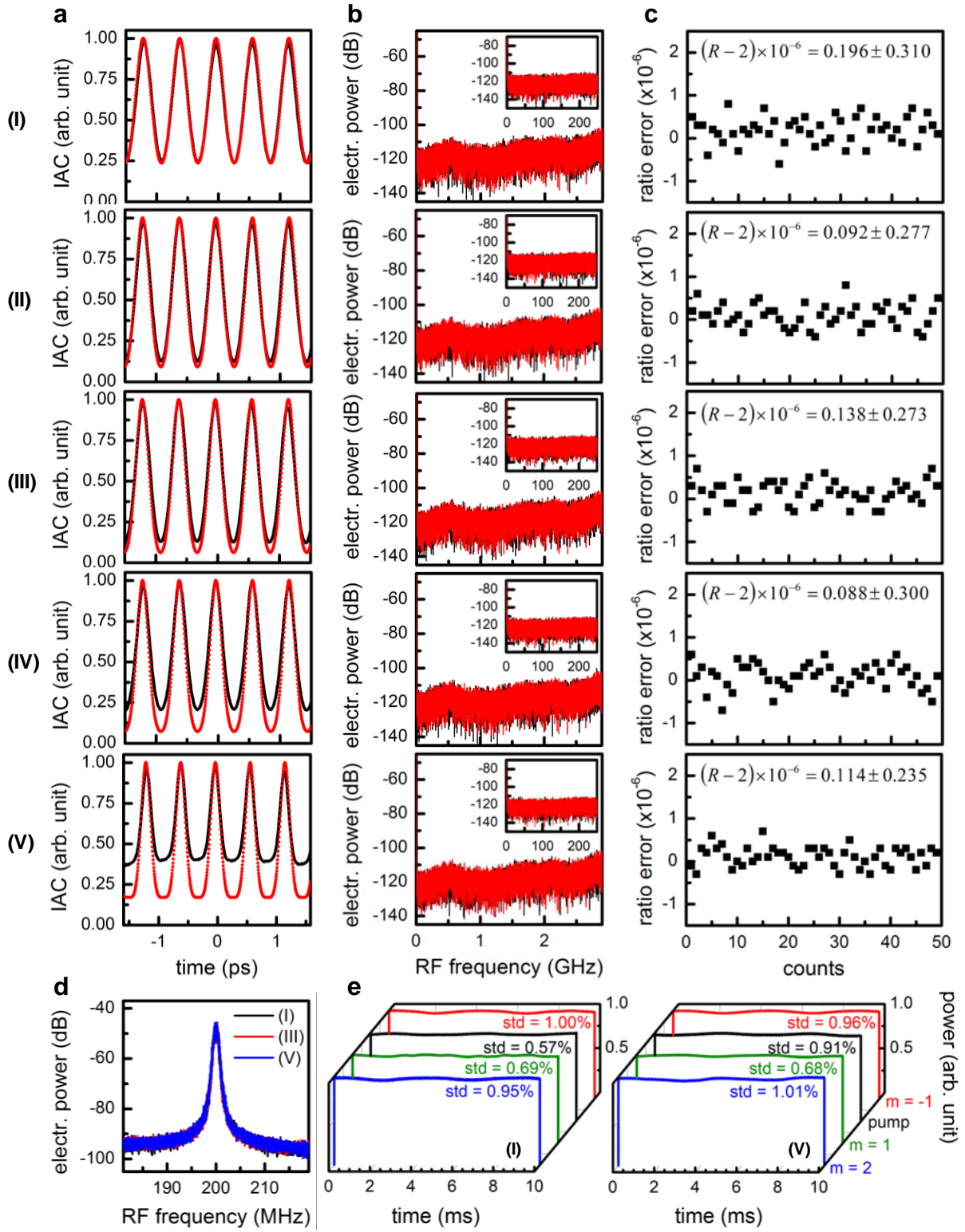

**Figure 3.** (**a**) IAC traces of the Turing rolls at different stages. The red dashed lines are the ideal traces calculated from the spectra, while the black curves are the measured traces. As the sidebands grow, the deviations between the measured and calculated IAC traces also increase. (**b**) To examine the emergence of incommensurate sub-combs, RF amplitude noise spectra of the Turing rolls (black curve) along with the detector background (red curve) are measured up to 3 GHz, six times the cavity linewidth. No apparent amplitude noise is observed, verifying the existence of the commensurate sub-combs. Inset: zoom-in RF amplitude noise spectra up to 250 MHz, at the instrumentation detection noise floor. (**c**) To probe the equidistance of the Turing rolls, the beat notes between the three Turing roll sidebands (pump, $m = 1$, and $m = 2$) and the adjacent fiber laser frequency comb lines are measured and the ratio errors are presented. The small deviation from the ideal ratio $R$ of 2 [$R$ defined in the main text as $(\delta_2 - \delta_0)/(\delta_1 - \delta_0)$] verifies Turing pattern's excellent uniformity. The average non-uniformity is measured at 125 mHz, or $7.3 \times 10^{-14}$ when referenced to the Turing pattern repetition rate at 1.72 THz. (**d**) The linewidth of the $1^{st}$ sideband is measured at 500 kHz, limited by the coherence of the pump laser, by the self-heterodyne technique at different stages. No linewidth broadening is observed, independently confirming that the coherence of the Turing roll is maintained at all evolving stages. (**e**) Power fluctuation of individual sidebands at the stages I (left) and V (right) with a sampling rate of 250 MHz, ruling out the possibility of breathing solutions.

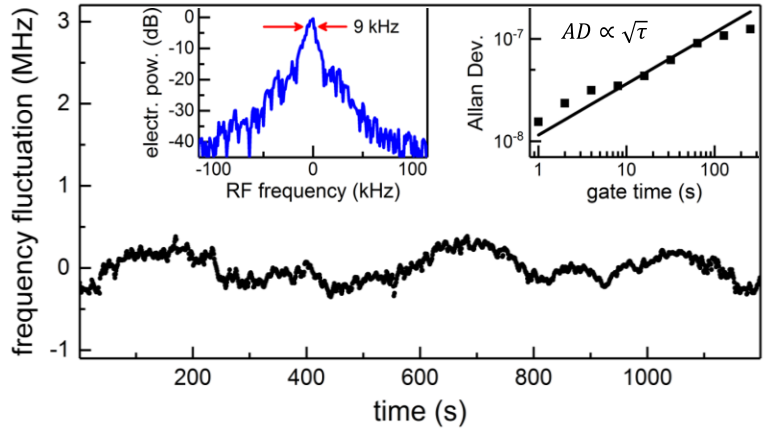

**Figure 4.** Long term frequency fluctuation of the 1$^{st}$ sideband with respect to the pump in the free-running mode, showing the repetition rate fluctuation of 160 kHz over 20 minutes. The left inset shows the linewidth of the beat note with a sweep time of 200 ms, measuring a narrow FWHM linewidth of 9 kHz; the right inset shows the frequency stability of the beat note, measuring an Allan deviation of $1.15 \times 10^{-8} \cdot \sqrt{\tau}$ when referenced to the THz carrier.

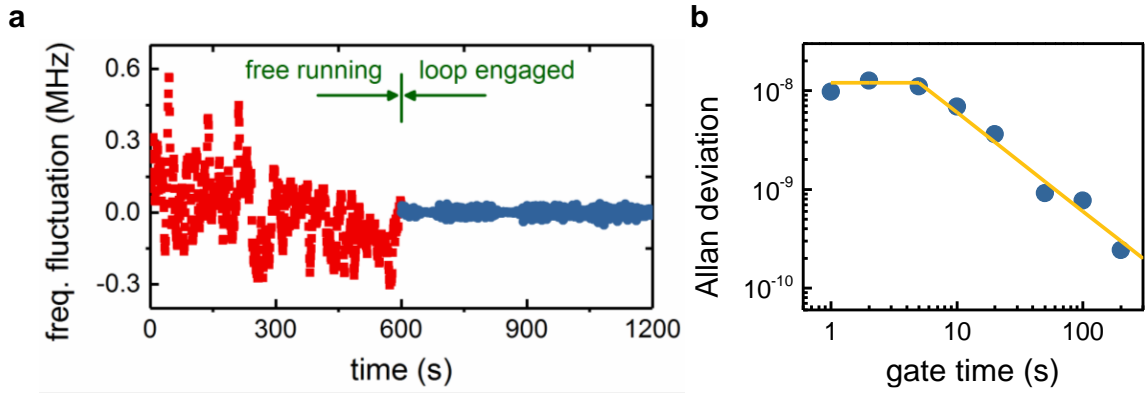

**Figure 5.** (**a**) Long term frequency fluctuation of the 1st sideband with respect to the pump without (red) and with (blue) engaging the computer-aided slow feedback control, counted with a gate time of 1 second. (**b**) Allan deviation of the stabilized THz frequency, measuring a plateau of $1.2\times10^{-8}$ for gate times shorter than 5 second and a roll-off of $6\times10^{-8}/\tau$ for longer gate times.

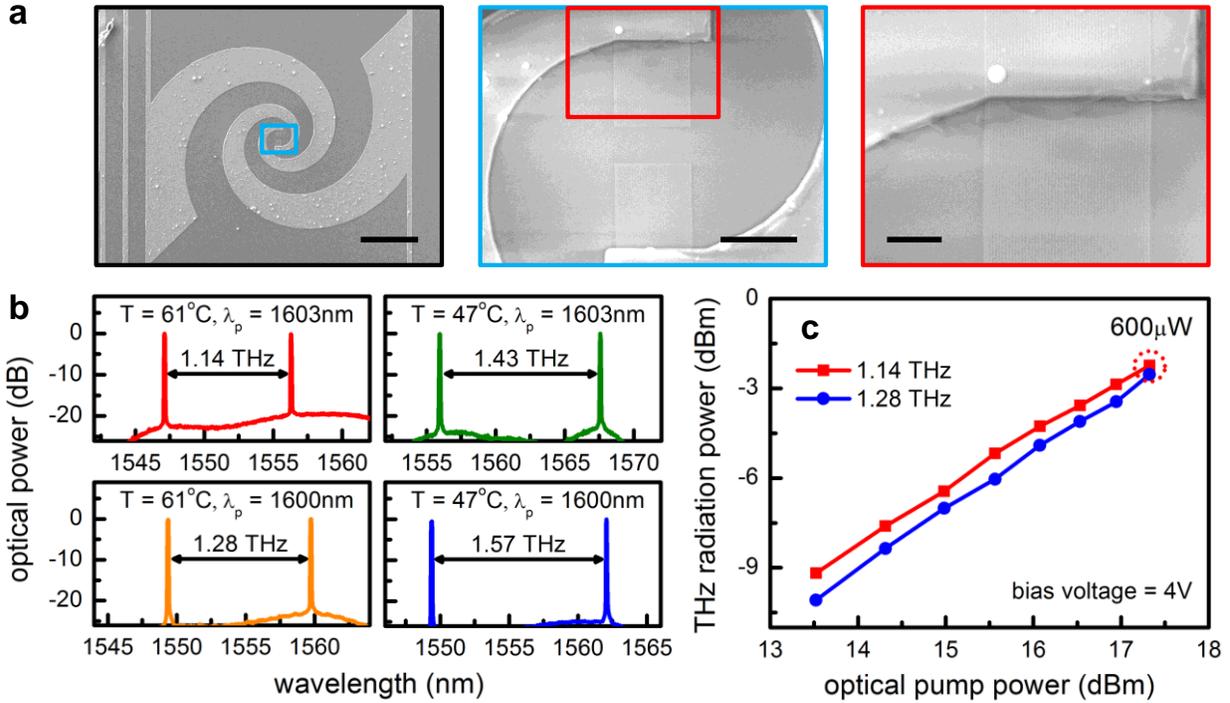

**Figure 6.** (**a**) Scanning electron micrographs of the fabricated plasmonic photomixer with a logarithmic spiral antenna integrated with plasmonic contact electrodes on an ErAs:InGaAs substrate. Scale bars from left to right: 100 μm, 10 μm, and 3 μm. (**b**) Turing roll repetition rate, and hence the generated THz frequency, can be tuned by changing the pump wavelength and the resonator temperature. (**c**) THz radiation power as a function of optical pump power. Power conversion efficiency of 1.1 % can be obtained with an optical pump power of 54 mW.